\documentclass{conm-p-l}
\usepackage{graphicx}

\newtheorem{theorem}{Theorem}[section]

\theoremstyle{definition}
\newtheorem{definition}[theorem]{Definition}
\newtheorem{example}[theorem]{Example}

\theoremstyle{remark}

\numberwithin{equation}{section}

\def\N{\mathbb N}
\def\Z{\mathbb Z}
\def\Q{\mathbb Q}
\def\S#1{S_{\!#1}}
\def\d{\partial}
\def\Nice{\mathrm{Nice}}
\def\lm{\mathrm{lm}}
\def\lc{\mathrm{lc}}

\def\bm#1{\mathchoice{\kern-.5pt\mathord{\text{\bfseries\itshape#1}}\kern+.25pt}
                      {\kern-.5pt\mathord{\text{\bfseries\itshape#1}}\kern+.25pt}
                      {\kern+.25pt\mathord{\text{\scriptsize\bfseries\itshape#1}}\kern+.5pt}
                      {\kern+.25pt\mathord{\text{\tiny\bfseries\itshape#1}}}\kern+.25pt} 
\def\Nice{\mathrm{Nice}}
\def\d{\partial}

\begin{document}

\title[An Algorithmic Proof of Stembridge's TSPP Theorem]
{Eliminating Human Insight: An Algorithmic Proof of Stembridge's TSPP Theorem}

\author{Christoph Koutschan}
\address{Research Institute for Symbolic Computation (RISC)\\
        Johannes Kepler University, A-4040 Linz, Austria}
\email{koutschan@risc.uni-linz.ac.at}
\thanks{supported by grant P20162 of the Austrian FWF}



\date{2009-05-27}

\begin{abstract}
  We present a new proof of Stembridge's theorem about the enumeration
  of totally symmetric plane partitions using the methodology
  suggested in the recent Koutschan-Kauers-Zeilberger semi-rigorous
  proof of the Andrews-Robbins $q$-TSPP conjecture. Our proof makes
  heavy use of computer algebra and is completely automatic.  We
  describe new methods that make the computations feasible in the
  first place.  The tantalizing aspect of this work is that the same
  methods can be applied to prove the $q$-TSPP conjecture (that is a
  $q$-analogue of Stembridge's theorem and open for more than 25
  years); the only hurdle here is still the computational complexity.
\end{abstract}

\maketitle

\section{Introduction}
The theorem (see Theorem \ref{tspp} below) that we want to address in
this paper is about the enumeration of \emph{totally symmetric plane
  partitions} (which we will abbreviate as TSPP, the definition is
given in Section~\ref{sectionTSPP}); it was first proven by John
Stembridge~\cite{Stembridge95}.  We will reprove the statement using
only computer algebra; this means that basically no human ingenuity
(from the mathematical point of view) is needed any more---once the
algorithmic method has been invented (see
Section~\ref{sectionMethod}).  But it is not as simple (otherwise this
paper would be needless): The computations that have to be performed
are very much involved and we were not able to do them with the known
methods. One option would be to wait for 20 years hoping that Moore's
law equips us with computers that are thousands of times faster than
the ones of nowadays and that can do the job easily. But we prefer a
second option, namely to think about how to make the problem feasible
for today's computers.  The main focus therefore is on presenting new
methods and algorithmic aspects that reduce the computational effort
drastically (Section~\ref{sectionAlgorithms}).  Our computations (for
the details read Section~\ref{sectionProof}) were performed in
\emph{Mathematica} using our newly developped package {\tt
  HolonomicFunctions}~\cite{Koutschan09}; this software will soon be available
on the RISC combinatorics software page 
\begin{center}
\fbox{http:/$\!$/www.risc.uni-linz.ac.at/research/combinat/software/}
\end{center}

Somehow, our results are a byproduct of a joint work with Doron
Zeilberger and Manuel Kauers~\cite{KauersKoutschanZeilberger09b} where
the long term goal is to apply the algorithmic proof method to a
$q$-analogue of Theorem~\ref{tspp} (see also
Section~\ref{sectionOutlook}).  The ordinary ($q=1$) case serves as a
proof-of-concept and to get a feeling for the complexity of the
underlying computations; hence it delivers valuable information
that go beyond the main topic of this paper.

Before we start we have to agree on some notation: We use the
symbol~$\S{n}$ to denote the shift operator, this means $\S{n}\bullet
f(n)=f(n+1)$ (in words ``$\S{n}$ applied to $f(n)$''). We use the
operator notation for expressing and manipulating recurrence
relations. For example, the Fibonacci recurrence $F_{n+2}=F_{n+1}+F_n$
translates to the operator~$\S{n}^2-\S{n}-1$.  When we do arithmetic
with operators we have to take into account the commutation rule
$\S{n}n=(n+1)\S{n}$, hence such operators can be viewed as elements in
a noncommutative polynomial ring in the indeterminates $n_1,\dots,n_d$
and $\S{n},\dots,\S{n_d}$.  Usually we will work with a structure
called Ore algebra, this means we consider an operator as a polynomial
in $\S{n_1},\dots,\S{n_d}$ with coefficients being rational functions
in $n_1,\dots,n_d$. Note that the noncommutativity now appears between
the indeterminates of the polynomial ring and the coefficients.  In
this context when speaking about the \emph{support} of an operator we
refer to the set of power products (monomials) in the $\S{n_i}$ whose
coefficient is nonzero.  For a given sequence we can consider the set
of all recurrences that this sequence fulfills; they form a left ideal
in the corresponding operator algebra.  We call it \emph{annihilating
  ideal} or in short \emph{annihilator} of the sequence. A sequence is
called $\d$-finite if there exists an annihilating ideal with the
property that only finitely many monomials can not be reduced by it,
in other words if the set of monomials that lie under the staircase of
a Gr\"obner basis of the ideal is finite. Together with the
appropriate set of initial values we refer to it as a
\emph{$\d$-finite description} of the sequence.

\section{Totally Symmetric Plane Partitions}\label{sectionTSPP}
In this section we want to give a short motivation of the combinatorial background of our problem.

\begin{definition}
A \emph{plane partition}~$\pi$ of some integer~$n$ is a two-dimensional array
\[
  \pi=(\pi_{ij}),\quad \pi_{ij}\in\N\mathrm{\ for\ integers\ }i,j\geq 1
\]
with finite sum $n=\sum_{i,j\geq 1} \pi_{ij}$
which is weakly decreasing in rows and columns, or more precisely
\[
  \pi_{i+1,j}\leq\pi_{ij}\quad\mathrm{and}\quad\pi_{i,j+1}\leq\pi_{ij}\quad\mathrm{for\ all\ }i,j\geq 1.
\]
\end{definition}

\begin{figure}
\begin{center}
\includegraphics[width=0.25\textwidth]{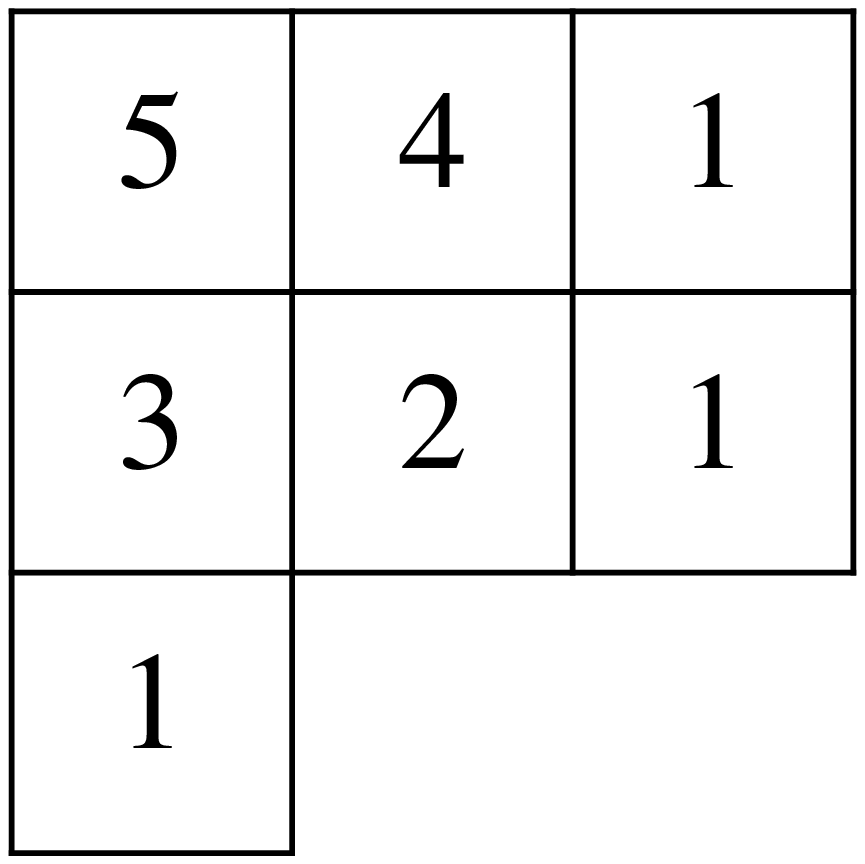}
\hskip 0.2\textwidth
\includegraphics[width=0.25\textwidth]{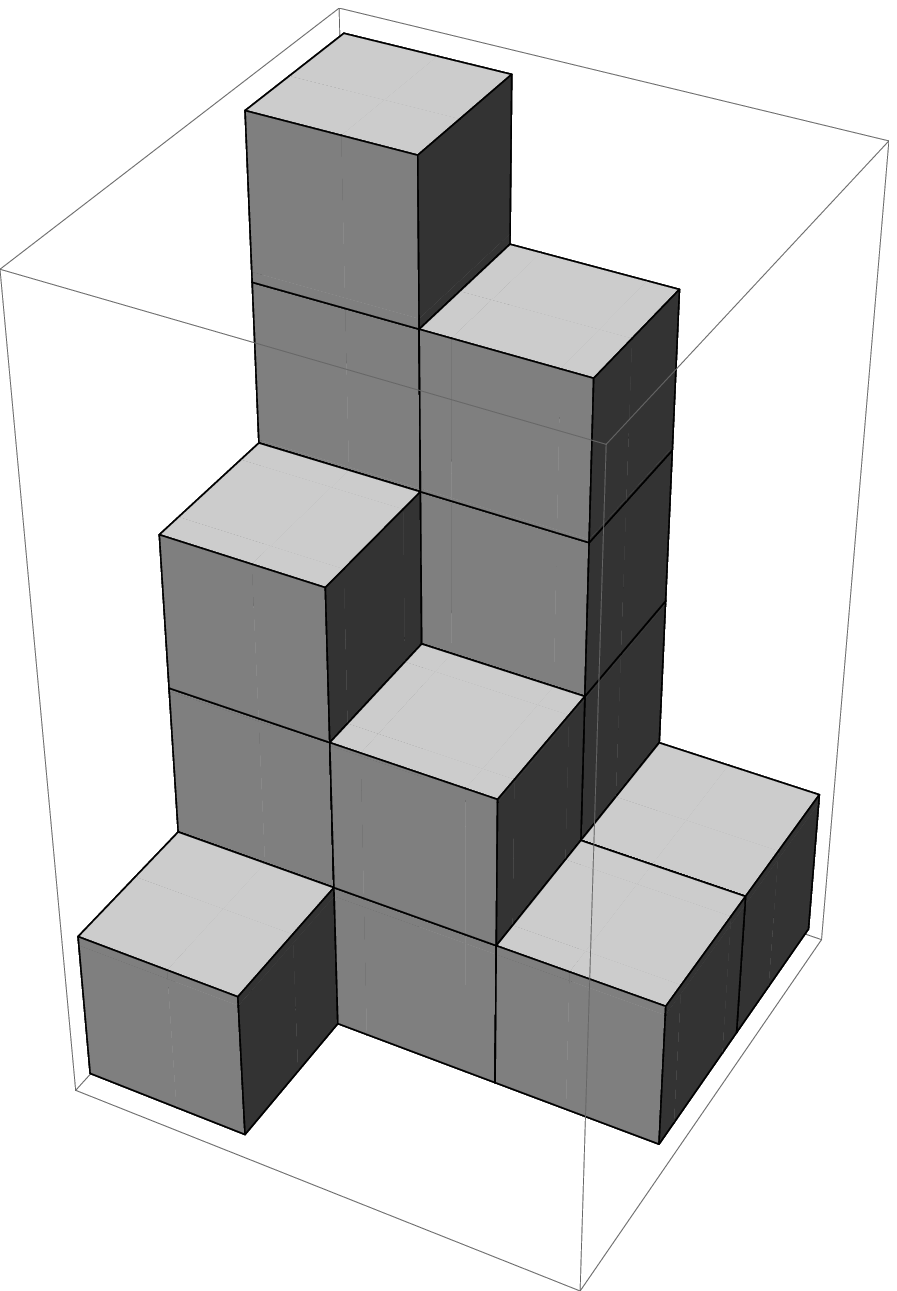}
\end{center}
\caption{A plane partition of $n=17$}
\label{figpp}
\end{figure}

Note that this definition implies that only finitely many
entries~$\pi_{ij}$ can be nonzero.  To each plane partition we can
draw its 3D~Ferrers diagram by stacking~$\pi_{ij}$ unit cubes on top
of the location $(i,j)$.  Each unit cube can be addressed by its
location $(i,j,k)$ in 3D~coordinates.  A 3D~Ferrers diagram is a
justified structure in the sense that if the position $(i,j,k)$ is
occupied then so are all positions $(i',j',k')$ with $i'\leq i$,
$j'\leq j$, and $k'\leq k$.  Figure~\ref{figpp} shows an example of a
plane partition together with its 3D~Ferrers diagram. We are now going
to define TSPPs, the objects of interest.

\begin{definition}
A plane partition is \emph{totally symmetric} iff
whenever the position $(i,j,k)$ in its 3D Ferrers diagram is occupied 
(in other words $\pi_{ij} \geq k$),
it follows that all its permutations 
$\{(i,k,j),(j,i,k),(j,k,i),(k,i,j),(k,j,i)\}$ are
also occupied. 
\end{definition}

Now Stembridge's theorem~\cite{Stembridge95} can be easily stated:
\begin{theorem}\label{tspp}
The number of totally symmetric plane partitions 
whose 3D~Ferrers diagram is contained in the cube
$[0,n]^3$ is given by the nice product-formula
\begin{equation}\label{tsppformula}
  \prod_{ 1 \leq i \leq j \leq k \leq n} \frac{i+j+k-1}{i+j+k-2}.
\end{equation}
\end{theorem}

\begin{figure}
\begin{center}
\includegraphics[width=0.75\textwidth]{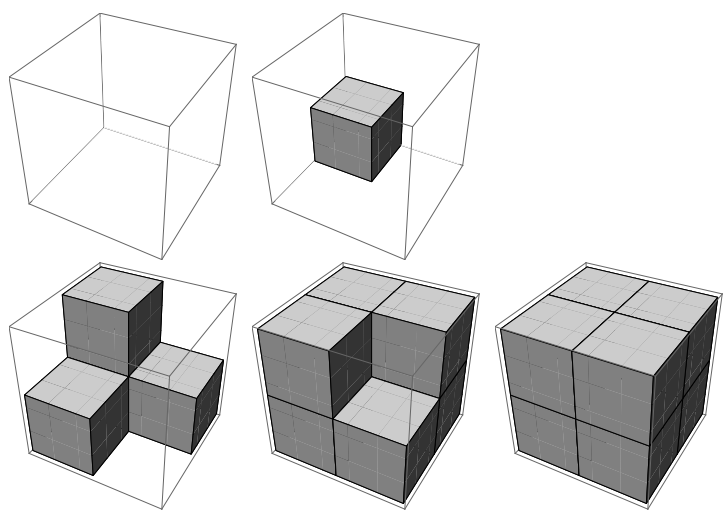}
\end{center}
\caption{All TSPPs that fit into the cube $[0,2]^2$}
\label{figtspp2}
\end{figure}

\begin{example} 
We are considering the case $n=2$: Formula \eqref{tsppformula} tells us that there should be
\[
  \prod_{ 1 \leq i \leq j \leq k \leq 2} \frac{i+j+k-1}{i+j+k-2}=
  \frac{2}{1}\cdot\frac{3}{2}\cdot\frac{4}{3}\cdot\frac{5}{4}=5
\]
TSPPs inside the cube~$[0,2]^3$ which is confirmed by the enumeration
given in Figure~\ref{figtspp2}.
\end{example}

As others that proved the TSPP formula before us we will make use of a
result by Soichi Okada \cite{Okada89} that reduces the proof of
Theorem~\ref{tspp} to a determinant evaluation:
\begin{theorem}
The enumeration formula \eqref{tsppformula} for TSPPs is correct if
and only if the determinant evaluation
\begin{equation}\label{determinant}
  \det\,(a(i,j))_{1 \leq i,j \leq n}=
  \prod_{ 1 \leq i \leq j \leq k \leq n} \left(\frac{i+j+k-1}{i+j+k-2}\right)^2
\end{equation}
holds, where the entries in the matrix are given by
\begin{equation}\label{aij}
  a(i,j)={i+j-2\choose i-1}+{i+j-1\choose i}+2\delta(i,j)-\delta(i,j+1).
\end{equation}
In the above, $\delta(i,j)$ denotes the Kronecker delta.
\end{theorem}

Ten years after Stembridge's proof, George Andrews, Peter Paule, and
Carsten Schneider \cite{AndrewsPauleSchneider05} came up with a
computer-assisted proof.  They transformed the problem into the task
to verify a couple of hypergeometric multiple-sum identities (which
they could do by the computer).  This problem transformation however
required human insight.  We claim to have the first ``human-free''
computer proof of Stembridge's theorem that is completely algorithmic
and does not require any human insight into the problem.  Moreover our
method generalizes immediately to the $q$-case which is not so obvious
to achieve in the approach presented in \cite{AndrewsPauleSchneider05}.

\section{Proof method for determinant evaluations}\label{sectionMethod}
Doron Zeilberger \cite{Zeilberger07} proposes a method for
completely automatic and rigorous proofs of determinant evaluations
that fit into a certain class.  For the sake of self-containedness
this section gives a short summary how the method works. 
It addresses the problem: For all $n \geq 0$ prove that
\[
  \det (a(i,j))_{1 \leq i,j \leq n} = \Nice(n),
\]
for some explicitly given expressions $a(i,j)$ and $\Nice(n)$.  What
you have to do is the following: Pull out of the hat another discrete
function $B(n,j)$ (this looks a little bit like magic for now---we
will make this step more explicit in the next section) and check the
identities
\begin{eqnarray}
\sum_{j=1}^n B(n,j)a(i,j)&=&0 \quad\mathrm{for\ }1 \leq i <n, \quad i,n\in\N,\label{soichi}\\
B(n,n)&=&1 \quad\mathrm{for\ all\ }n\geq 1,\quad n\in\N.\label{normalization}
\end{eqnarray}

Then by uniqueness, it follows that $B(n,j)$ equals the cofactor of
the $(n,j)$~entry of the $n \times n$ determinant (i.e. the minor with
the last row and the $j$th column removed, this means we expand the
determinant with respect to the last row using Laplace's formula),
divided by the $(n-1)\times(n-1)$ determinant. In other words we
normalized in a way such that the last entry $B(n,n)$ is 1. Or, to
make the argument even more explicit: What happens if we replace the
last row of the matrix by any of the other rows? Clearly then the
determinant will be zero; and nothing else is expressed in
equation~\eqref{soichi}.

Finally one has to verify the identity
\begin{equation}\label{okada}
\sum_{j=1}^{n} B(n,j) a(n,j)= \frac{\Nice(n)}{\Nice(n-1)} \quad\mathrm{for\ all\ }n\geq 1,\quad n\in\N.
\end{equation}
If the suggested function~$B(n,j)$ does satisfy all these identities
then the determinant identity follows immediately as a consequence.

\section{The algorithms}\label{sectionAlgorithms}
We now explain how the existing algorithms (in short) as well as our
approach (in more detail) find a recurrence for some definite sum.  In
order to keep the descriptions simple and concrete we consider a sum
of the form
\[
  \sum_{j=1}^nf(n,j)
\]
as it appears in \eqref{okada} (everything generalizes to instances
with more parameters in the summand as it is the case in
\eqref{soichi}).  We give some indications why the existing algorithms
fail to work in practice; all these statements refer to \eqref{okada}
but apply in a similar fashion to \eqref{soichi} as well.

\subsection{Some unsuccessful tries}
There are several methods in the literature how to algorithmically
prove identities like \eqref{soichi} and \eqref{okada}.  The first one
traces back to Doron Zeilberger's seminal paper \cite{Zeilberger90} and
he later named it the \emph{slow algorithm}. The idea is to find a
recurrence operator in the annihilating ideal of the summand that does not
contain the summation variable in its coefficients; such a relation
can always be rewritten in the form
\[
  P(n,\S{n})+(\S{j}-1)Q(n,\S{j},\S{n})
\]
and we call~$P$ the \emph{principal part} and~$Q$ the \emph{delta
  part}. Such a \emph{telescoping relation} encodes that~$P$ is a
recurrence for the sum (depending on the summand and the delta part we
might have to add an inhomogeneous part to this recurrence).  The
elimination can be performed by a Gr\"obner basis computation with
appropriate term order. In order to get a handle on the variable $j$
we have to consider the recurrences as polynomials in $j$, $\S{j}$,
and $\S{n}$ with coefficients in~$\Q(n)$ (for efficiency reasons this
is preferable compared to viewing the recurrences as polynomials in
all 4 indeterminates with coefficients in~$\Q$).  We tried this
approach but it seems to be hopeless: The variable~$j$ that we would
like to eliminate occurs in the annihilating relations for the summand
$B(n,j)a'(n,j)$ with degrees between 24 and 30.  When we follow the
intermediate results of the Gr\"obner basis computation
we observe that none of the elements that were added to the basis
because some S-polynomial did not reduce to zero has a degree in $j$
lower than 23 (we aborted the computation after more than 48 hours).
Additionally the coefficients grow rapidly and it seems very likely
that we run out of memory before coming to an end.

The second option that we can try is often referred to as Takayama's
algorithm~\cite{Takayama90b}. In fact, we would like to
apply a variant of Takayama's original algorithm that was proposed by
Chyzak and Salvy~\cite{ChyzakSalvy98}. Concerning speed this algorithm
is much superior to the elimination algorithm described above: It
computes only the principal part~$P$ of some telescoping operator
\begin{equation}\label{telescoping}
  P(n,\S{n})+(\S{j}-1)Q(j,n,\S{j},\S{n}).
\end{equation}
When we sum over natural boundaries we need not to know about the
delta part~$Q$. This is for example the case when the summand has only
finite support (which is the case in our application). Also this
algorithm boils down to an elimination problem which, as before, seems
to be unsolvable with today's computers: We now can lower the degree
of~$j$ to 18, but the intermediate results consume already about 12GB
of memory (after 48 hours).

The third option is Chyzak's algorithm \cite{Chyzak00} for $\d$-finite
functions: It finds a relation of the form \eqref{telescoping} by
making an ansatz for~$P$ and~$Q$; the input recurrences are
interpreted as polynomials in~$\S{j}$ and~$\S{n}$ with coefficients
being rational functions in~$j$ and~$n$. It uses the fact that the
support of~$Q$ can be restricted to the monomials under the stairs of
the input annihilator and it loops over the order of~$P$.  Because of
the multiplication of~$Q$ by $\S{j}-1$ we end up in solving a coupled
linear system of difference equations for the unknown coefficients
of~$Q$.  Due to the size of the input, we did not succeed in
uncoupling this system, and even if we can do this step, it remains to
solve a presumably huge (concerning the size of the coefficients as
well as the order) scalar difference equation.

\subsection{A successful approach}
The basic idea of what we propose is very simple: We also start with an
ansatz in order to find a telescoping operator. But in contrast to Chyzak's
algorithm we avoid the expensive uncoupling and solving of difference equations.
The difference is that we start with a polynomial ansatz in~$j$ up to some
degree:
\begin{equation}\label{ansatz}
  \underbrace{\sum_{i=0}^Ic_i(n)\S{n}^i}_{=\>P(n,\S{n})}\quad
  +\quad(\S{j}-1)\cdot\underbrace{\sum_{k=0}^K\sum_{l=0}^L\sum_{m=0}^Md_{k,l,m}(n)j^k\S{j}^l\S{n}^m}_{=\>Q(j,n,\S{j},\S{n})}.
\end{equation}
The unknown functions~$c_i$ and~$d_{k,l,m}$ to solve for are rational
functions in~$n$ and they can be computed using pure linear
algebra. Recall that in Chyzak's algorithm we have to solve for
rational functions in~$n$ and~$j$ which causes the system to be
coupled. The prize that we pay is that the shape of the ansatz is not
at all clear from a priori: The order of the principal part, the
degree bound for the variable~$j$ and the support of the delta part
need to be fixed, whereas in Chyzak's algorithm we have to loop only
over the order of the principal part. Our approach is similar to the
generalization of Sister Celine Fasenmyer's technique that is used in
Wegschaider's {\tt MultiSum} package \cite{Wegschaider97} (which can
deal with multiple sums but only with hypergeometric summands). We
proceed by reducing the ansatz with a Gr\"obner basis of the given
annihilating left ideal for the summand, obtaining a normal form
representation of the ansatz. Since we wish this relation to be in the
ideal, the normal form has to be identically zero. Equating the
coefficients of the normal form to zero and performing coefficient comparison
with respect to~$j$ delivers a linear system for the unknowns that has
to be solved over~$\Q(n)$.

Trying out for which choice of $I,K,L,M$ the ansatz delivers a
solution can be a time-consuming tedious task. Additionally, once a
solution is found it still can happen that it does not fit to our
needs: It can well happen that all~$c_i$ are zero in which case the
result is useless.  Hence the question is: Can we simplify the search
for a good ansatz, for example, by using homomorphic images?  Clearly
we can reduce the size of the coefficients by computing modulo a prime
number (we may assume that the input operators have coefficients in
$\Z[j,n]$, otherwise we can clear denominators). But in practice this
does not reduce the computational complexity too much---still we have
bivariate polynomials that can grow dramatically during the reduction
process. For sure we can not get rid of the variable~$j$ since it is
needed later for the coefficient comparison. It is also true that we
can not just plug in some concrete integer for~$n$: We would lose the
feature of noncommutativity that~$n$ shares with~$\S{n}$ (recall that
$\S{n}n=(n+1)\S{n}$, but $\S{n}7=7\S{n}$ for example). And the
noncommutativity plays a crucial role during the reduction process, in
the sense that omitting it we get a wrong result. Let's have a closer
look what happens and recall how the normal form computation works:

\begin{center}\begin{tabular}{|p{1mm}l|}\hline
\multicolumn{2}{|l|}{\rule{0mm}{4mm}Algorithm: Normal form computation}\\[0.5ex] \hline
\multicolumn{2}{|l|}{\rule{0mm}{4mm}{\bf Input:} an operator $p$ and a Gr\"obner basis $G=\{g_1,\dots,g_m\}$}\\
\multicolumn{2}{|l|}{{\bf Output:} normal form of $p$ modulo the left ideal $\langle G\rangle$}\\[0.5ex] \hline
\multicolumn{2}{|l|}{\rule{0mm}{4mm}while exists $1\leq i\leq m$ such that $\lm(g_i)\,\mid\,\lm(p)$}\\
& $g:=(\lm(p)/\lm(g_i))\cdot g_i$\\
& $p:=p-(\lc(p)/\lc(g))\cdot g$\\
\multicolumn{2}{|l|}{end while}\\
\multicolumn{2}{|l|}{return $p$}\\[0.5ex] \hline
\end{tabular}\end{center}
where $\lm$ and $\lc$ refer to the leading monomial and the leading
coefficient of an operator respectively.

Note that we do the multiplication of the polynomial that we want to
reduce with in two steps: First multiply by the appropriate power
product of shift operators (line 2), and second adjust the leading
coefficient (line 3). The reason is because the first step usually
will change the leading coefficient. Note also that~$p$ is never
multiplied by anything.  This gives rise to a modular version of the
normal form computation that does respect the noncommutativity.

\begin{center}\begin{tabular}{|p{1mm}l|}\hline
\multicolumn{2}{|l|}{\rule{0mm}{4mm}Algorithm: Modular normal form computation}\\[0.5ex] \hline
\multicolumn{2}{|l|}{\rule{0mm}{4mm}{\bf Input:} an operator $p$ and a Gr\"obner basis $G=\{g_1,\dots,g_m\}$}\\
\multicolumn{2}{|l|}{{\bf Output:} modular normal form of $p$ modulo the left ideal $\langle G\rangle$}\\[0.5ex] \hline
\multicolumn{2}{|l|}{\rule{0mm}{4mm}while exists $1\leq i\leq m$ such that $\lm(g_i)\,\mid\,\lm(p)$}\\
& $g:=h((\lm(p)/\lm(g_i))\cdot g_i)$\\
& $p:=p-(\lc(p)/\lc(g))\cdot g$\\
\multicolumn{2}{|l|}{end while}\\
\multicolumn{2}{|l|}{return $p$}\\[0.5ex] \hline
\end{tabular}\end{center}
where $h$ is an insertion homomorphism, in our example
$h:\Q(j,n)\to\Q(j)$, $h(f(j,n))\mapsto f(j,n_0)$ for some
$n_0\in\N$. Thus most of the computations are done modulo the
polynomial $n-n_0$ and the coefficient growth is moderate compared to
before (univariate vs. bivariate).

Before starting the nonmodular computation we make the ansatz as small
as possible by leaving away all unknowns that are~0 in the modular
solution. With very high probability they will be~0 in the final
solution too---in the opposite case we will realize this unlikely
event since then the system will turn out to be unsolvable.  In
\cite{Wegschaider97} a method called Verbaeten's completion is used in
order to recognize superfluous terms in the ansatz a priori.  We were
thinking about a generalization of that, but since the modular
computation is negligibly short compared to the rest, we don't expect
to gain much and do not investigate this idea further.

Other optimizations concern the way how the reduction is performed.
With a big ansatz that involves hundreds of unknowns (as it will be
the case in our work) it is nearly impossible to do it in the naive
way.  The only possibility to achieve the result at reasonable cost is
to consider each monomial in the support of the ansatz
separately. After having computed the normal forms of all these
monomials we can combine them in order to obtain the normal form of
the ansatz. Last but not least it pays off to make use of the
previously computed normal forms. This means that we sort the
monomials that we would like to reduce according to the term order in
which the Gr\"obner basis is given. Then for each monomial we have to
perform one reduction step and then plug in the normal forms that we
have already (since all monomials that occur in the support after the
reduction step are smaller with respect to the chosen term order).

\section{The computer proof}\label{sectionProof}
We are now going to give the details of our computer proof of Theorem
\ref{tspp} following the lines described in the previous section.

\subsection{Get an annihilating ideal}\label{getann}
The first thing we have to do according to Zeilberger's algorithmic
proof technique is to resolve the magic step that we have left as a
black box so far, namely ``to pull out of the hat'' the
sequence~$B(n,j)$ for which we have to verify the identities
\eqref{soichi} -- \eqref{okada}.  Note that we are able, using the
definition of what~$B(n,j)$ is supposed to be (namely a certain minor
in a determinant expansion), to compute the values of~$B(n,j)$ for
small concrete integers~$n$ and~$j$.  This data allows us (by plugging
it into an appropriate ansatz and solving the resulting linear system
of equations) to find recurrence relations for~$B(n,j)$ that will hold
for all values of~$n$ and~$j$ with a very high probability.  We call
this method \emph{guessing}; it has been executed by Manuel Kauers who
used his highly optimized software {\tt Guess.m} \cite{Kauers09}. More
details about this part of the proof can be found in
\cite{KauersKoutschanZeilberger09b}.  The result of the guessing were
65 recurrences, their total size being about 5MB.

Many of these recurrences are redundant and it is desirable to have a
unique description of the object in question that additionally is as
small as possible (in a certain metric).  To this end we compute a
Gr\"obner basis of the left ideal that is generated by the 65
recurrences. The computation was executed by the author's
noncommutative Gr\"obner basis implementation which is part of the
package {\tt HolonomicFunctions}.  The Gr\"obner basis consists of 5
polynomials (their total size being about 1.6MB).  Their leading
monomials $\S{j}^4,\S{j}^3\S{n},\S{j}^2\S{n}^2,\S{j}\S{n}^3,\S{n}^4$
form a staircase of regular shape.  This means that we should take 10
initial values into account (they correspond to the monomials under
the staircase).

In addition, we have now verified that all the 65 recurrences are
consistent. Hence they are all describing the same object. But since
we want to have a rigorous proof we have to admit at this point that
what we have found so far (that is a $\d$-finite description of some
bivariate sequence---let's call it $B'(n,j)$) does not prove anything
yet. We have to show that this~$B'(n,j)$ is identical to the
sequence~$B(n,j)$ defined by \eqref{soichi} and \eqref{normalization}.
Finally we have to show that identity~\eqref{okada} indeed holds.

\subsection{Avoid singularities}
\begin{figure}
\begin{center}
\includegraphics[width=0.4\textwidth]{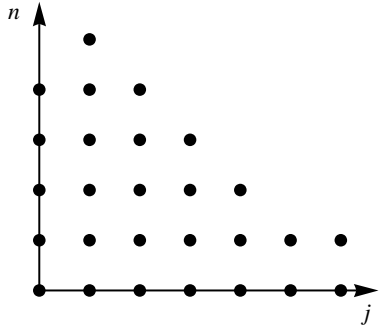}
\end{center}
\caption{The points for which the initial values of the sequence $B(n,j)$ have to be given 
because the recurrences do not apply.}
\label{figsing}
\end{figure}

Before we start to prove the relevant identities there is one subtle
point that, aiming at a fully rigorous proof, we should not omit: the
question of singularities in the $\d$-finite description of~$B'(n,j)$.
Recall that in the univariate case when we deal with a P-finite
recurrence, we have to regard the zeros of the leading coefficient
and in case that they introduce singularities in the range where we
would like to apply the recurrence, we have to separately specify the
values of the sequence at these points. Similarly in the bivariate
case: We have to check whether there are points in~$\N^2$ where none
of the recurrences can be applied because the leading term
vanishes. For all points that lie in the area $(4,4)+\N^2$ we may
apply any of the recurrences, hence we have to look for common
nonnegative integer solutions of all their leading coefficients. A
Gr\"obner basis computation reveals that everything goes well: From
the first element of the Gr\"obner basis
\[
  (n-3)^2 (n-2) (n-1)^2 (2 n-3)^2 (2 n-1) (j+n-1) (j+n)
\]
we can read off the solutions $(0,0)$, $(1,0)$, and $(0,1)$ 
(which are also solutions of the remaining polynomials but since
they are lying under the stairs they are of no interest).
Further we have to address the cases $n=1,2,3$.
Plugging these into the remaining polynomials we obtain further common
solutions: $(1,1)$, $(2,1)$, $(2,2)$, $(3,2)$, and $(3,3)$. 
But all of them are outside of $(4,4)+\N^2$ so we need not to care.
It remains to look at the lines $j=0,1,2,3$ and the lines $n=0,1,2,3$
(we omit the details here). Summarizing, the points for which initial
values have to be given (either because they are under the stairs or because
of singularities) are
\[
  \begin{array}{l}
  \{(j,n)\mid 0\leq j\leq6\land 0\leq n\leq1\}\cup\{(j,2)\mid 0\leq j\leq4\}\cup\{(j,3)\mid 0\leq j\leq3\}\cup\\
  \{(j,4)\mid 0\leq j\leq2\}\cup\{(1,5)\}.
  \end{array}
\]
They are depicted in Figure~\ref{figsing}.

\subsection{The second identity}
The simplest of the three identities to prove is
\eqref{normalization}.  From the $\d$-finite description of~$B'(n,j)$
we can compute a recurrence for the diagonal~$B'(n,n)$ by the closure
property ``substitution''. {\tt HolonomicFunctions} delivers a
recurrence of order~7 in a couple of minutes. Reducing this recurrence
with the ideal generated by $\S{n}-1$ (which annihilates 1) gives~0; 
hence it is a left multiple of the recurrence for the right hand
side. We should not forget to have a look on the leading coefficient
in order to make sure that we don't run into singularities:
\[
  256 (2 n+3) (2 n+5) (2 n+7) (2 n+9) (2 n+11)^2 (2 n+13)^2 p_1 p_2
\]
where $p_1$ and $p_2$ are irreducible polynomials in $n$ of degree 4 and 12 respectively.
Comparing initial values (which of course match due to our definition) establishes
identity \eqref{normalization}. 

\subsection{The third identity}
In order to prove \eqref{okada} we first rewrite it slightly.
Using the definition of the matrix entries $a(n,j)$ we obtain
for the left hand side
\[
  \sum_{j=1}^n B(n,j)\underbrace{\left({n+j-2\choose n-1}+{n+j-1\choose n}\right)}_{=:a'(n,j)}
  \> + 2B(n,n) - B(n,n-1)
\]
and the right hand side simplifies to
\[
  \frac{\Nice(n)}{\Nice(n-1)}=
  \frac{\prod_{ 1 \leq i \leq j \leq k \leq n} \left(\frac{i+j+k-1}{i+j+k-2}\right)^2}
       {\prod_{ 1 \leq i \leq j \leq k \leq n-1} \left(\frac{i+j+k-1}{i+j+k-2}\right)^2}=
  \frac{4^{1-n} (3 n-1)^2 (2 n)_{n-1}^2}{(3 n-2)^2 (n/2)_{n-1}^2}.
\]
Note that $a'(n,j)=\frac{2n+j-1}{n+j-1}{n+j-1\choose j-1}$ is a
hypergeometric expression in both variables $j$ and $n$.  A
$\d$-finite description of the summand can be computed with {\tt
  HolonomicFunctions} from the annihilator of $B(n,j)$ by closure
property.  We found by means of modular computations that the ansatz
\eqref{ansatz} with $I=7$, $K=5$, and the support of $Q$ being the
power products~$\S{j}^l\S{n}^m$ with $l+m\leq 7$ delivers a solution
with nontrivial principal part. After omitting the 0-components of
this solution, we ended up with an ansatz containing 126 unknowns.
For computing the final solution we used again homomorphic images and
rational reconstruction. Still it was quite some effort to compute the
solution (it consists of rational functions in~$n$ with degrees up to
382 in the numerators and denominators).  The total size of the
telescoping relation becomes smaller when we reduce the delta part to
normal form (then obtaining an operator of the form that Chyzak's
algorithm delivers). Finally the result takes about 5 MB of memory. We
counterchecked its correctness by reducing the relation with the
annihilator of $B(n,j)a'(n,j)$ and obtained~0 as expected.

We have now a recurrence for the sum but we need to to cover the whole
left hand side. A recurrence for $B(n,n-1)$ is easily obtained with
our package performing the substitution $j\to n-1$, and $B(n,n)=1$ as
shown before. The closure property ``sum of $\d$-finite functions''
delivers a recurrence of order~10. On the right hand side we have a
$\d$-finite expression for which our package automatically computes an
annihilating operator. This operator is a right divisor of the one
that annihilates the left hand side. By comparing 10 initial values
and verifying that the leading coefficients of the recurrences do not
have singularities among the positive integers, we have established
identity~\eqref{okada}.

\subsection{The first identity}
With the same notation as before we reformulate identity~\eqref{soichi} as
\[
  \sum_{j=1}^nB(n,j)a'(i,j)=B(n,i-1)-2B(n,i).
\]
The hard part again is to do the sum on the left hand side.  Since two
parameters $i$ and $n$ are involved and remain after the summation,
one annihilating operator does not suffice.  We decided to search for
two operators with leading monomials being pure powers of $\S{i}$ and
$\S{n}$ respectively. Although this is far away from being a Gr\"obner
basis, it is nevertheless a complete description of the object
(together with sufficiently (but still finitely) many initial
values). We obtained these two relations in a similar way as in the
previous section, but the computational effort was even bigger
(more than 500 hours of computation time were needed).  The first
telescoping operator is about 200 MB big and the support of its
principal part is \setlength{\arraycolsep}{0mm}
\[
 \begin{array}{ll}
 \{&\S{i}^5, \S{i}^4\S{n}, \S{i}^3\S{n}^2, \S{i}^2\S{n}^3, \S{i}\S{n}^4, \S{i}^4, \S{i}^3\S{n},
 \S{i}^2\S{n}^2, \S{i}\S{n}^3, \\
 &\S{i}^3, \S{i}^2\S{n}, \S{i}\S{n}^2, \S{n}^3, \S{i}^2, \S{i}\S{n}, \S{n}^2, \S{i}, \S{n}, 1\}.
 \end{array}
\]
The second one is of size 700 MB and the support of its principal part is 
\[
  \{\S{n}^5, \S{i}^4, \S{i}^3\S{n}, \S{i}^2\S{n}^2, \S{i}\S{n}^3, \S{n}^4, \S{i}^3, \S{i}^2\S{n},
  \S{i}\S{n}^2, \S{n}^3, \S{i}^2, \S{i}\S{n}, \S{n}^2, \S{i}, \S{n}, 1\}.
\]
Again we can independently from their derivation check their
correctness by reducing them with the annihilator of $B(n,j)a'(i,j)$:
both give~0.

Let's now address the right hand side: From the Gr\"obner basis for
$B(n,j)$ that we computed in Section \ref{getann} one immediately gets
the annihilator for $B(n,i-1)$ by replacing $\S{j}$ by $\S{i}$ and by
substituting $j\to i-1$ in the coefficients. We now could apply the
closure property ``sum of $\d$-finite functions'' but we can do
better: Since the right hand side can be written as $(1-2\S{i})\bullet
B(n,i-1)$ we can use the closure property ``application of an
operator'' and obtain a Gr\"obner basis which has even less monomials
under the stairs than the input, namely 8. The opposite we expect to
happen when using ``sum'': usually there the dimension grows but never
can shrink.  It is now a relatively simple task to verify that the two
principal parts that were computed for the left hand side are elements
of the annihilating ideal of the right hand side (both reductions
give~0).

The initial value question needs some special attention here since we
want the identity to hold only for $i<n$; hence we can not simply look
at the initial values in the square $[0,4]^2$. Instead we compare the
initial values in a trapezoid-shaped area which allows us to compute
all values below the diagonal. Since all these initial values match
for the left hand and right hand side we have the proof that the
identity holds for all $i<n$. Looking at the leading coefficients of
the two principal parts we find that they contain the factors $5+i-n$
and $5-i+n$ respectively.  This means that both operators can not be
used to compute values on the diagonal which is a strong indication
that the identity does not hold there: Indeed, identity~\eqref{soichi}
is wrong for $n=i$ because in this case we get~\eqref{okada}.

\section{Outlook}\label{sectionOutlook}
As we have demonstrated Zeilberger's methodology is completely
algorithmic and does not need human intervention. This fact makes it
possible to apply it to other problems (of the same class) without
further thinking.  Just feed the data into the computer! The $q$-TSPP
enumeration formula
\[
  \prod_{1 \leq i \leq j \leq k \leq n} \frac{1-q^{i+j+k-1}}{1-q^{i+j+k-2}}
\]
has been conjectured independently by George Andrews and Dave Robbins
in the early 1980s.  This conjecture is still open and one of the most
intriguing problems in enumerative combinatorics. The method as well
as our improvements can be applied one-to-one to that problem (also a
$q$-analogue of Okada's result exists). Unfortunately, due to the
additional indeterminate $q$ the complexity of the computations is
increased considerably which prevents us from proving it right
away. But we are working on that\dots

\medskip\noindent\textbf{Acknowledgements.}
I would like to thank Doron Zeilberger for attentively following
my efforts and providing me with helpful hints. Furthermore
he was the person who came up with the idea to attack TSPP in the way we did.
Special thanks go to my colleague Manuel Kauers with whom
I had lots of fruitful discussions during this work and who
performed the guessing part in Section~\ref{getann}. He also
provided me with his valuable knowledge and software on how to efficiently
solve linear systems using homomorphic images.

\bibliographystyle{amsplain}


\providecommand{\bysame}{\leavevmode\hbox to3em{\hrulefill}\thinspace}
\providecommand{\MR}{\relax\ifhmode\unskip\space\fi MR }
\providecommand{\MRhref}[2]{%
  \href{http://www.ams.org/mathscinet-getitem?mr=#1}{#2}
}
\providecommand{\href}[2]{#2}

\end{document}